\documentclass{appolb}
\usepackage[a4paper, left= 0.9in, right=0.9in, top=0.6in, bottom=0.6in]{geometry}
\usepackage{amsmath}
\usepackage{graphicx} 
\usepackage{authblk}

\begin{document}
\title{Effects of self-gravity of the accretion disk around rapidly rotating black hole in long GRBs
}
\author{Ishika Palit$^{1}$ \thanks{Email: palit@cft.edu.pl}, Agnieszka Janiuk$^{1}$
\address{$^{1}$ Center for Theoretical Physics PAS, Al. Lotnikow 32/46, 02-668 Warsaw, Poland}
\\
{and Petra Sukova$^{2}$}
\address{$^{2}$ Astronomical Institute CAS, Bo\v{c}n\'{i} II 1401, CZ-141 00 Prague, Czech Republic}
}
\maketitle

\begin{abstract}
We prescribe a method to study the effects of self gravity of accretion disk around a black hole associated with long Gamma Ray Bursts (GRBs) in an evolving background Kerr metric. This is an extension to our previous work where we had presented possible constraints for the final masses and spins of these astrophysical black holes. Incorporating the self force of the accreting cloud around the black hole is a very important aspect due to the transient nature of the event, in which a huge amount of mass is accreted and changes the fundamental black hole parameters i.e. its mass and spin, during the process. Understanding of GRBs engine is important because they are possible sources of high-energy particles and gravitational waves as most of the energy released from the dynamical evolution is in the form of gravitational radiation. Here we describe the analytical framework we developed to employ in our numerical model. The numerical studies are planned for the future work.
\end{abstract}

\section*{ Introduction}
Gamma Ray Bursts (GRBs) are highly energetic explosions in the universe releasing a tremendous amount of energy. Based on observation, these energetic events can be classified into two different types: short-duration bursts with a duration of less than 2 seconds and long duration bursts lasting more than 2 seconds \cite{Kouvelioutou1996}. The most conceded theory to describe the long-duration gamma ray bursts is the collapsar model where a very massive, rapidly rotating star collapses down to its iron core and forms a black hole \cite{woosley1993, woosley2012, piran}. Not all the stars produce long GRBs as they collapse and the distinguishing reason is the rotation of the star because a disk may not form in case of insufficient angular momentum \cite{janiuk2008, janiuk2008b}. The in-falling matter from the collapsed star, falling onto this black hole forms an accretion disk around it which can launch a pair of relativistic jets (beams of ionized matter) along the rotational axis of the progenitor star. These jets further scrape off through the stellar surface and produce emission in gamma rays.
\\
In the case of long GRBs, the process of collapse inevitably involves feeding the new black hole with mass and angular momentum. Thus the computation of a GRB engine in a dynamically evolving space-time metric is very important specifically due to the transient nature of the event, in which a huge amount of mass is accreted and changes the fundamental black hole parameters i.e. its mass and spin, during the process. The inclusion of the gravitational force of accreting disk in the study of GRBs is crucial because this creates local instabilities if the mass density of the accreting cloud becomes comparable to $ M/R^{3} $ , where M and R are the mass of the central object and the disk radius, respectively. Self gravity in accretion theory limits the angular momentum of the disk as well as affects the evolution of black hole mass. 

\section*{Analytical model}
To incorporate the effects of self gravity of the disk, we chose the Teukolsky equation.  This equation describes gravitational, electro-magnetic, scalar and neutrino field perturbations of a rotating Kerr black hole \cite{Teukolsky1972}. The global vacuum solution of Teukolsky equation is given by CCK method \cite{d,a,b,c} which reconstructs the metric perturbation and shows that only perturbations of the mass and angular momentum ($\delta M$ and $\delta J$) are to be concluded within the Kerr metric. Further Maarten van de Meent \cite{Maarten2017} shows that the perturbation due to a particle for any bound orbit around a Kerr black hole described by CCK metric only affects the Kerr parameters describing the mass and angular momentum of the black hole in the Kerr metric 'outside' the particle's orbit and vanishes 'inside' the orbit.
\\
The most important thing to specify here is that we are dealing with fluids but not particles in our simulation. That is the reason for incorporating the perturbation in our models, calculated as volume integrals of the corresponding components of stress-energy tensor containing the mass-energy and angular momentum of the gas.
We consider orbits around the black hole with radius equal to each grid point of our simulation. The perturbative terms here are the only function of radius and time : $\delta M (t,r)$ and $\delta J(t,r)$. The update of mass and spin of the black hole at any given time, t and point, r due to perturbation is calculated considering the volume inside the sphere of radius r. So the total mass of the black hole at a time is the sum of black hole initial mass, accreted amount of mass through the horizon, and perturbed mass inside the volume at that specified point, r at that time.
\begin{equation}
M(t,r) = M_{0} + \Delta M (t) + \delta M(t,r)
\end{equation}
where $M_{0}$ is the initial mass of the black hole. $\Delta M(t)$ is the amount of mass energy accreted through the horizon $(r = r_{horizon})$ i.e the inner boundary so far:
\begin{equation}
\Delta M(t) = \int_{0}^{t} \int_{0}^{2\pi} \int_{0}^{\pi} \sqrt{-g} T^{r}_{t} d\theta d\phi dt^{'}
\end{equation}
$\delta M(t,r)$ is the actual amount of mass energy of the gas inside the
sphere from the horizon up to radius r of the particle and the volume integral gives: 
\begin{equation}
\delta M(t,r) = \int_{r_{horizon}}^{r} \int_{0}^{2\pi} \int_{0}^{\pi} \sqrt{-g} T^{t}_{t} d\theta d\phi dr^{'}
\end{equation}
Similarly the angular momentum can be calculated :
\begin{equation}
J(t,r) = J_{0} + \Delta J (t) + \delta J(t,r)
\end{equation}
where $J_{0}$ is the initial angular momentum of the black hole,
\\
$\Delta J(t)$ is the amount of angular momentum accreted through the horizon so far :
\begin{equation}
\Delta J(t) = \int_{0}^{t} \int_{0}^{2\pi} \int_{0}^{\pi} \sqrt{-g} T^{r}_{\phi} d\theta d\phi dt^{'}
\end{equation}
$\delta J(t,r)$ is the actual amount of angular momentum of the gas inside the sphere from the horizon up to radius r of the particle and the volume integral gives : 
\begin{equation}
\delta J(t,r) = \int_{r_{horizon}}^{r}  \int_{0}^{2\pi} \int_{0}^{\pi} \sqrt{-g} T^{t}_{\phi} d\theta d\phi dr^{'}
\end{equation} 

Our simulation considers a non-magnetized, quasi-spherical flow. We plan to follow its evolution numerically. We will make an array of $\delta M$ and $\delta J$  which are a function of radius and time. At each time, these arrays will consist of values equal to the number of our grid points in r direction, calculated for the volume inside each radius or grid point. Thus each point will have different amounts of perturbed mass.
\\
Calculation of the perturbative terms, $\delta M$ and $\delta J$ will be further used in the calculation of the metric coefficients to update mass, M and spin, $a$ of the black hole where  $a(t,r) = \frac{J(t,r)} {M(t,r)}$. During the simulations, the units for mass, length and time is given by the initial mass of the black hole $M_{0}$.

\section*{Numerical Model}
To construct a physical model of the accretion disk, we need to solve the GRMHD equations further supplemented by the equation of state (EoS) of the matter. In the scenario of GRBs, the EoS is complex. Under the conditions of extremely high densities and temperatures, the nuclear reactions have to be taken into account. All these physical complexities: magnetic fields, general relativity, nuclear reactions, pose a challenge to any kind of numerical scheme. There are few studies dealing with the micro-physics of gamma ray bursts \cite{janiuk2017, janiuk2019}. 
\\
In our current approach, we will neglect both the micro-physics and neutrino transport in the simulation, and we will be using the adiabatic equation of state. This assumption will help the simulation to go faster and efficiently. It is beyond scope of our code for now to deal with the detailed microphysics of the flow in addition to the dynamical evolution of the perturbative metric update.
\\
We are also considering a non-magnetized flow. Ignoring the aspect of the magnetic field in the case of GRBs leads to a very simplified case but the inclusion of magnetized flow in such updating Kerr metric is very much numerically challenging.
\\
We will implement this CCK method described above in our version of GRMHD numerical code- HARM (High Accuracy Relativistic Magnetohydrodynamic) \cite{Gammie2004} (our version:\cite{janiuk18}). It is a conservative, shock capturing scheme, for evolving the equations of GRMHD  (General Relativistic Magneto HydroDynamics). The code solves numerically the continuity equation, the four-momentum-energy conservation equation, and induction equation in GR framework. The form of integrated equation in the code is :
\begin{equation}
\delta U_{t}(P) = - \delta_{i} F^{i}(P) + S(P)
\end{equation} 
where U is a vector of conserved variables, such as particle number density, energy and momentum, $F^{i}$ are the fluxes in finite control volume, and S is a vector of source terms. The vector P is composed of primitive variables, such as rest-mass density, internal energy density, velocity components, and magnetic field components, which are interpolated to model the flow within zones. HARM solves GRMHD equations in the modified version of Kerr-Schild coordinate system (KS) rather than Boyer-Lindquist coordinates thus matter can accrete smoothly through the horizon and evolution of the flow can be followed properly without reaching any coordinate singularity.

\section*{Summary}
Our previous study \cite{janiuk2018} speculates on the possible constraints for the final masses and spins of these astrophysical black holes. The study shows how much the evolution of flow is sensitive to the changes in the space-time metric (see Fig.[1]). It is a strong point in favor of the inclusion of self gravity effects of the disk on the metric in order to investigate such dynamical evolution. We expect to see more mass and spin growth of the black hole after incorporating the perturbation effects of the accreting disk into the updating metric.
\begin{figure}[ht]
\includegraphics[width = 1.0\textwidth]{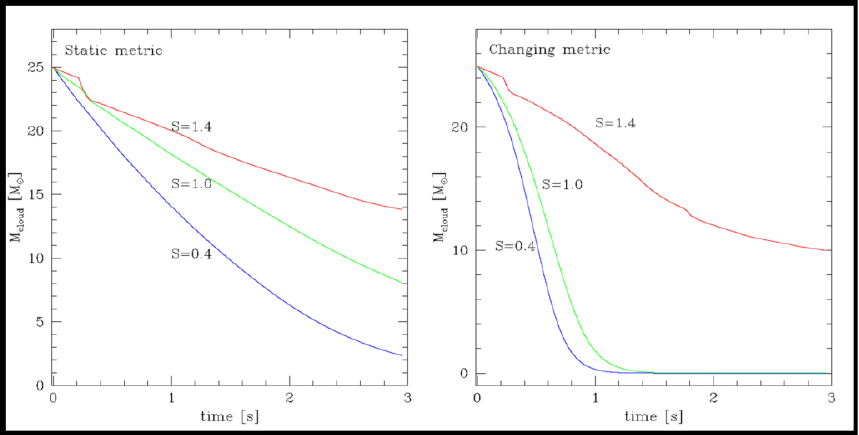}\\
\caption{Mass of the accreting cloud, contained within 1000 gravitational radii ($r_{g}$), as a function of time. Three solid lines represent the rotation with critical angular momentum at $6 r_{g}$(green), and for 0.4 $l_{critical}$ (blue), and for 1.4  $l_{critical}$ (red). Figure adopted from \cite{janiuk2018}.}
\label{fig3}
\end{figure} 
The importance of having different $\delta M$ and $\delta J$ at each grid point in r direction at each time affects the metric coefficients which are sensitive to mass and spin update. We plan to follow the dynamical evolution of such a system and study the mass growth, spin evolution and disk structure in this scenario. We are also going to investigate the variation in mass accretion rate and the following power density spectra from the corresponding light curve.
\\
In our model, we consider that the black hole already been formed in the center and the surrounding matter is feeding the black hole via accretion. We are not going to study the formation of this black hole by the collapse of the massive star by solving Einstein field equations with the total matter field i.e the stress-energy tensor including fluid and the radiation part as well. There is no such detailed model present to date as per my knowledge following the evolution from the very collapse of the star to the growth of the black hole in a complete general relativistic framework with detailed microphysics of the flow. It is very complicated to investigate all the aspects associated with the phenomenon in a single model (see this review about GRB progenitors : \cite{fryer}).
\\
We are going to implement this above described analytical framework in our numerical study to get a more clear and realistic analysis of mass growth, spin evolution and disk structure in the collapsar scenario. Our study would join two important aspects of the collapsar model for GRBs studied so far separately and provide a more unified, pragmatic and detailed model for the very first time. 
\section*{Acknowledgement}
We thank Vladmir Karas and Vojtěch Witzany from Astronomical Institute CAS, for helpful discussions. 
We also acknowledge support from the Interdisciplinary Center for Mathematical Modeling of the Warsaw University, through the computational grant Gb79-9, and the PL-Grid computational resources through the grant grb2. PS is supported from Grant No. GACR-17-06962Y from Czech Science Foundation.

 

\end{document}